\newcommand{\beq}{\begin{equation}}
\newcommand{\eeq}{\end{equation}}

\newcommand{\bear}{\begin{eqnarray}}
\newcommand{\eear}{\end{eqnarray}}

\newcommand{\tn}{\textnormal}

\documentclass[review]{elsart5p}

\usepackage{graphicx}
\usepackage{enumerate}
\usepackage{pifont}
\usepackage{amssymb}
\usepackage{bbding}
\usepackage{lineno}
\usepackage{amsmath}
\usepackage{amssymb}
\usepackage{dsfont}

\begin{document}

\begin{frontmatter}

\title{Modelling the behaviour of microbulk Micromegas in Xenon/trimethylamine gas}
\author[1,2] {E. Ruiz-Choliz},
\author[1,2,3]{D. Gonz\'alez-D\'iaz\corauthref{aut1}},
\corauth[aut1]{Corresponding author.}\ead{diegogon@cern.ch}
\author[1,2] {A. Diago},
\author[1,2] {J. Castel},
\author[1,2] {T. Dafni},
\author[1,2] {D. C. Herrera},
\author[1,2] {F. J. Iguaz},
\author[1,2] {I. G. Irastorza},
\author[1,2] {G. Luz\'on},
\author[1,2] {H. Mirallas},
\author[19]   {\"{O}. \c{S}ahin},
\author[19,3] {R. Veenhof}

\address[1]{Laboratorio de F\'isica Nuclear y Astropart\'iculas, Universidad de Zaragoza, Zaragoza, Spain}
\address[2]{Laboratorio Subterr\'aneo de Canfranc, Canfranc, Spain}
\address[3]{CERN, Geneva, Switzerland}
\address[19]{Department of Physics, Uluda\u{g} University, Bursa, Turkey}

\begin{abstract}
We model the response of a state of the art micro-hole single-stage charge amplification device (`microbulk' Micromegas) in a gaseous atmosphere consisting of Xenon/trimethylamine at various concentrations and pressures. The amplifying structure, made with photo-lithographic techniques similar to those followed in the fabrication of gas electron multipliers (GEMs), consisted of a 100~$\mu$m-side equilateral-triangle pattern with 50~$\mu$m-diameter holes placed at its vertexes. Once the primary electrons are guided into the holes by virtue of an optimized field configuration, avalanches develop along the 50~$\mu$m-height channels etched out of the original doubly copper-clad polyimide foil. In order to properly account for the strong field gradients at the holes' entrance as well as for the fluctuations of the avalanche process (that ultimately determine the achievable energy resolution), we abandoned the hydrodynamic framework, resorting to a purely microscopic description of the electron trajectories as obtained from elementary cross-sections. We show that achieving a satisfactory description needs additional assumptions about atom-molecule (Penning) transfer reactions and charge recombination to be made.
\end{abstract}

\end{frontmatter}


\section{Introduction}
\label{intro}

Microbulk Micromegas, introduced in \cite{muBulk}, are a new generation of charge-amplifying Micromegas devices \cite{MMyannis} capable of delivering, on a selected mm$^2$ spot, intrinsic energy resolutions down to 10.5\%@5.9keV (Neon/i-C$_{4}$H$_{10}$), 11.2\%@5.9keV (Argon/i-C$_{4}$H$_{10}$) and 7.3\%@22keV (Xenon/TMA) at atmospheric pressure \cite{muBulk,PacoMeas,XeTMADiana}. A similar hierarchy (He/Ne/Ar/Xe) has been experimentally observed for micro-meshes under sub-${100~\mu}$m$^2$ irradiation in \cite{Zerguerras}, and it can be linked to the efficiency of the avalanche multiplication process,
as theoretically anticipated \cite{RobSta}.


It is indeed remarkable that the performance of microbulk readouts deteriorates only slightly on real-sized experimental systems, showing 8.5\%@30keV (Xenon/TMA) on 700 cm$^2$ \cite{XeTMADiego}, and 14.7\%@5.9keV (Argon/i-C$_{4}$H$_{10}$) on 15 cm$^2$ \cite{XaviRadio}. 
Furthermore, the 730 normal liter TPC in \cite{XeTMADiego} was operated steadily under irradiation from a 15~kBq $\gamma$-ray source for more than 100 live days. A modest $\times 2$ reduction of the maximum working gain, relative to values obtained earlier for small 3.5 cm$^2$ wafers in \cite{XeTMADiana} was observed. Given the large number of holes typically employed ($\sim 10^8/$m$^2$), such good scaling properties require a great accuracy on the hole diameter and gas gap, at the $\%$-level or below.

Besides offering an accurate mechanical construction, capable of delivering a good energy resolution and stability on large areas, additional benefits of the microbulk technique are the low material budget and relatively high radiopurity \cite{HectorRadio}. Their overall performance makes them attractive for the construction of large time projection chambers (TPCs) for experiments on `Rare Event Searches' under a broad range of admixtures and pressures. For some applications, chiefly solar axion detection, radioactivity levels of 10$^{-7}$/s/cm$^2$/keV \cite{XaviRadio} could be achieved in the sub-10~keV energy region for underground operation.

Microbulk Micromegas are presently used in CAST \cite{XaviRadio} and n-TOF \cite{nTOF} experiments, and its application is foreseen or considered in a new generation of experiments involving axions (IAXO \cite{IAXO}), dark matter (TREX-DM \cite{TREX}) and $\beta\beta0\nu$ decay \cite{HPXeChina}. The latter, naturally focusing on high-pressure $^{136}$Xe, builds on the pioneering works on charge-readouts performed by the Gotthard collaboration \cite{Gotthard} and more recently in \cite{XeTMADiego}, and will try to challenge the state of the art of $\beta\beta0\nu$ detection in gas, based on light-readout and presently led by the NEXT experiment \cite{NEXT1, NEXT2}.

Outside the ideal conditions experimentally realized in \cite{Zerguerras}, for the practical operation of Micromegas readouts an interplay exists between geometrical tolerance, optimal gap and pressure \cite{YannisComp,NarrowGap}; between charge loss in the drift/conversion region and field focusing \cite{XeTMADiana,HerreraReco}; and between high signal to noise ratio and low photon-feedback \cite{Balan, XeTMADiego}.
Ideally, considering the readout alone, one would wish to operate at the lowest possible drift field and highest possible gain before the probability of feedback (or breakdown), recombination or attachment become important, as well as using a gas mixture and geometry that show a minimal sensitivity to the geometrical tolerances, \cite{YannisComp}. For microbulk Micromegas, these aspects have not been discussed in full yet.

We focus in this work on the modelling of microbulk Micromegas operated under Xenon and trimethylamine (TMA) for various admixtures and pressures. We benefited from the detailed experimental survey performed in \cite{XeTMADiana}, and the microscopic code Garfield++ \cite{Garfield} in conjunction with the field solver Elmer \cite{Elmer} and Gmsh \cite{gmesh}. Considering that the Penning and recombination characteristics of the mixture might be at the core of an intriguing idea for directional Dark Matter detection \cite{Dave,Jin} and provided the only existing estimate for those properties is approximate \cite{DiegoProc}, we attempted to extend the results from the microscopic electron tracking to include those additional effects.




\section{Simulation}

We followed for the simulation of the Micromegas response an approach based on the newly developed Garfield++ \cite{Garfield}.
The framework allows to track electrons by considering their
elementary interactions with atoms and molecules, through the electron cross-sections tabulated
in Magboltz \cite{Magboltz}. It provides a more generic way of evaluating electron trajectories than the implicit
 thermalization ansatz of the hydrodinamic formalism (also available in Garfield) while at the
same time it allows accessing the fluctuations in the avalanche multiplication process. The latter
are essential for a proper description of the energy resolution in gaseous detectors, that is attempted here. Ability for
describing the avalanche fluctuations \cite{Zerguerras, RobSta}, and electron transparency through micro-meshes \cite{RobTransp} has been demonstrated with this approach earlier. Nonetheless, the simulation framework does not contain at the moment libraries for the computation of electron-ion recombination or atom-atom/molecule-atom transfer reactions, that are of relevance for the present work, and that have been hence evaluated externally.

For the 3D-field calculation we relied on the open source packages Elmer and Gmsh, that are interfaced with Garfield++.
The mesh was generated through tetrahedrons on a elementary cell, that was then replicated to reproduce a volume large enough to avoid fringe effects.

%
%

\section{Micromegas geometry and electron transmission}

\subsection{Geometry}
\begin{figure}[h!!!]
 \centering
 \includegraphics*[width=\linewidth]{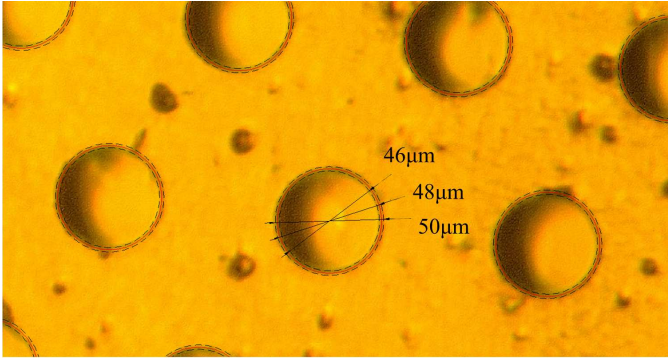}
 \caption{Microscope image of the surface of the microbulk Micromegas whose performance has been simulated in this work. The surface characteristics are determined by the gold coating. The dashed circles represent the biggest and smallest circles compatible with the hole foot-print. The (red) continuous circle has the average diameter of both, yielding an estimate of $\phi=48\mu$m$ \pm 2\mu$m.}
 \label{Geometry}
 \end{figure}

\begin{figure}[h!!!]
 \centering
 \includegraphics*[width=5.5cm]{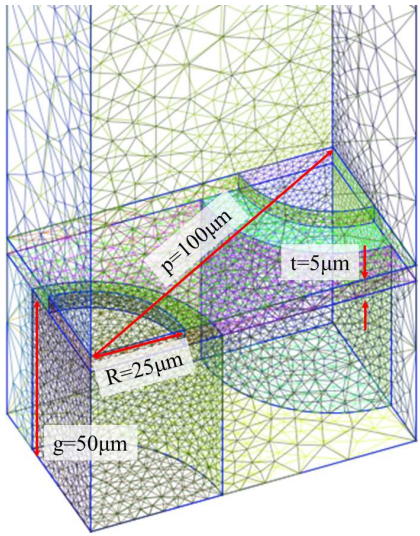}
 \caption{3D mesh used for the field calculation (in color online).}
 \label{GeometryElmer}
 \end{figure}

The geometry of the Micromegas chosen for this work is the one used in \cite{XeTMADiana}, an upper view of which is shown in Fig. \ref{Geometry}, taken with a microscope. Its pattern consists of holes of diameter $\phi= 48~{\mu}$m$~ \pm~ 2~ {\mu}$m$(\tn{sys}) ~\pm ~2~ {\mu}$m$ (\tn{sta})$ arranged as the vertexes of equilateral triangles of sides $p=100~{\mu}$m$~ \pm ~2~{\mu}$m$(\tn{sta})$. The systematic uncertainty in the hole diameter comes from its slight eccentricity, so the given range actually comprises the smallest and largest circles compatible with the hole boundaries (dashed circles in Fig. \ref{Geometry}). The statistical uncertainty comes from the measuring procedure. We have chosen in the following $\phi=50~\mu$m and $p=100~\mu$m, that were in fact the target values provided to the photo-lithographic mask. The 3D electric field modelling was performed in the elementary cell shown in Fig. \ref{GeometryElmer}, that extends along the drift/conversion region over 1~cm height. The thickness of the copper cladding (nano-coated with gold) was 5~$\mu$m and the multiplication channel along the polyimide core was emptied in simulation by 20 additional $\mu$m. The actual assumption is not critical for the simulated hole gain (representing a 20\% effect at most) and is immaterial for the computation of the electron transmission probability. The idea here is to eliminate any charge being lost to the insulator through diffusion, an effect that would likely cause gain transients and field reconfiguration (see \cite{RobTrans} and references therein), and that is known to be small in microbulk Micromegas. The actual amount of polyimide that is chemically etched inwards from the channel walls is not precisely known. By replicating the field in the elementary cell, the simulation was extended to a region of approximately $10 \times 10$ holes, ensuring no charge loss outside the geometrical boundaries under the conditions studied.

\subsection{Relative fraction of collected charge}
During Micromegas operation, it is difficult to unambiguously isolate the fraction of charge directly gained in amplification from the one lost in the drift region or in the drift-to-hole electron transmission. An experimental determination of the optimal charge collection conditions for operation usually involves a gain scan as a function of the drift field (and normalized to the maximum gain obtained along the scan), hereafter dubbed $\mathcal{F}$. In simulation, on the other hand, the `true' electron transmission $\mathcal{T}_{MC}$ can be obtained easily by determining the fraction of primary electrons entering the holes of the structure (i.e., crossing a plane defined by the upper copper layer in Fig. \ref{GeometryElmer}).

Electrons were launched in simulation at 200 $\mu$m from the holes' plane, uniformly distributed on an elementary cell.
The results presented were found to be independent from the injection plane and from the particular distribution chosen (as long as the injection distance was greater than $200~\mu$m). Pre-amplification in the drift region was found to be negligible even in the presence of sizeable Penning transfer rates, so during the determination of the transmission curves this process was neglected. The finite element mesh density chosen in Fig. \ref{GeometryElmer} provided a good compromise in terms of computing time, with finer meshes not yielding any visible difference. The asymptotic behaviour of the electron transmission (before the onset of pre-amplification) agreed with the optical transparency of the Micromegas within 5\%, implying that field lines become largely perpendicular to the hole plane. Although realistic electron trajectories in gases involve an additional random motion, in-hole and out-of-hole diffusion statistically cancel out for such high electric field conditions.

\begin{figure}[h!!!]
\centering
\includegraphics*[width=7.5cm]{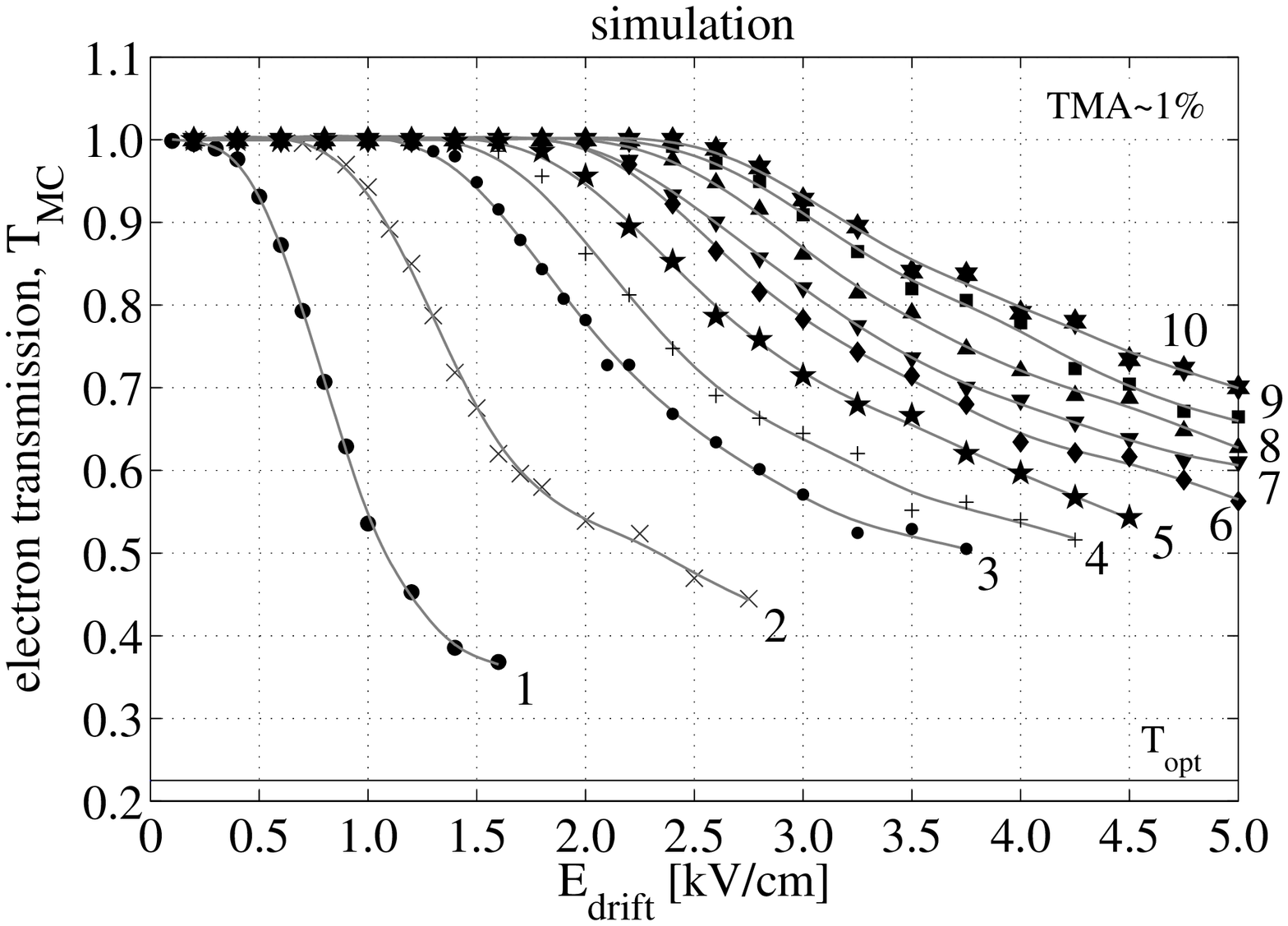}
\includegraphics*[width=7.5cm]{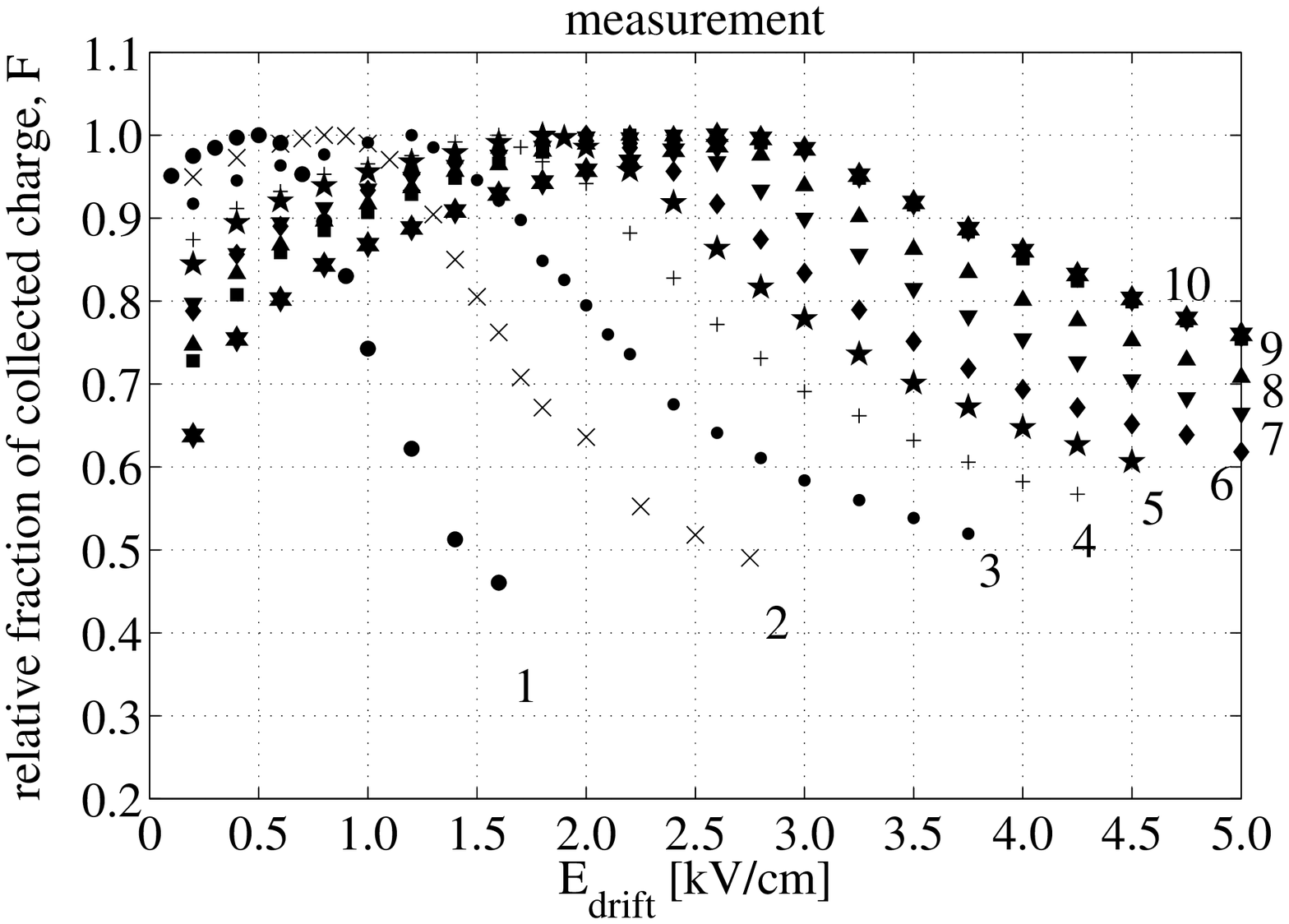}
\caption{Simulated electron transmission (up), and fraction of collected charge $\mathcal{F}$ determined experimentally. The labels indicate the operating pressure in bar. From 1 to 10~bar the TMA concentrations are, respectively: 0.90$\%$, 1.01$\%$, 1.15$\%$, 0.86$\%$, 0.86$\%$, 0.86$\%$, 0.79$\%$, 0.79$\%$, 0.79$\%$, 0.79$\%$. Simulation results are presented with super-imposed splines for guiding the eye. The horizontal line indicates the optical transparency.}
\label{SimulAndData}
\end{figure}

The simulated transmission and the relative charge fraction obtained experimentally are shown in Fig. \ref{SimulAndData} as a function of the drift field ($E_d$), for gains around 200-1000 and for a TMA concentration of $\sim1$\%. A $^{109}$Cd source was used for the measurements, and the analysis performed by assesing the variations of the $\varepsilon=22$~keV peak. First and foremost, it is apparent that this type of comparison provides an excellent cross-check tool for the assessment of the geometry and quality of the structure.\footnote{Prior to this work, and due to a miscommunication, the hole pitch was thought to be $115\mu$m, resulting in a gross disagreement.} Electron transmission is expected to depend on the geometry, electric field configuration and gas properties, the latter known to yield an adequate description of the drift velocity and diffusion coefficients in Xe-TMA \cite{XeTMADiegoOLD}. Excellent agreement had been reported earlier, e.g., in \cite{RobTransp}, for the transmission through standard micro-meshes.

Despite the good agreement in Fig. \ref{SimulAndData}, small deviations between measurements and simulation can still be found both at high and (specially) low fields. To the right of the maximum of each series, the measured transmission is systematically rightwards shifted, by about 10\%. The effect exceeds any systematic error stemming from the calibration of the power supply (3\%), drift distance (2\%), temperature during gas injection ($<3$\%), TMA concentration (10\%) and TMA inelastic cross-sections (20\%). Variations along these lines end short at explaining the observations, however when considering the assumed uncertainty in the reference geometry ($2{\mu}$m) this additional shift can be accommodated easily.

To the left of the maximum of each series in Fig. \ref{SimulAndData}, charge losses due to either attachment or diffusion outside the active volume could be a priori expected, translating into a dependence of the collected charge with the drift distance $z$. This seems to be incompatible, however, with the long electron life-times ($\sim5$ms) measured in the system \cite{HerreraReco, XeTMADiego}, implying charge losses below 0.1\% for the 1~cm drift employed. Additionally, the X-ray collimation was performed down to a few mm$^2$ at the center of a 3.5 cm diameter Micromegas sensor, un unbridgeable distance for the emerging photo-electron. Ballistic defect can be discarded down to a sub-$\%$ contribution, due to the $50\mu$s integration time and the known values for the diffusion, drift velocity and photo-electron cloud \cite{XeTMADiego}.\footnote{Since the spatial dimensions of the primary ionization cloud are compressed at high pressure and the drift velocity is largely a function of $E_d/P$, a pressure increase would reduce ballistic deficit and thus increase the collected charge for any given $E_d/P$, in stark contrast to the observed behaviour.}
The most obvious culprit of the decrease of the collected charge at low fields is the presence of losses due to charge recombination. This effect, studied systematically in \cite{HerreraReco}, has been recently reported in an ionization chamber measurement under 30~keV X-rays \cite{Yasu}, qualitatively agreeing with the behaviour reported here.

\begin{figure}[h!!!]
 \centering
 \includegraphics*[width=8cm]{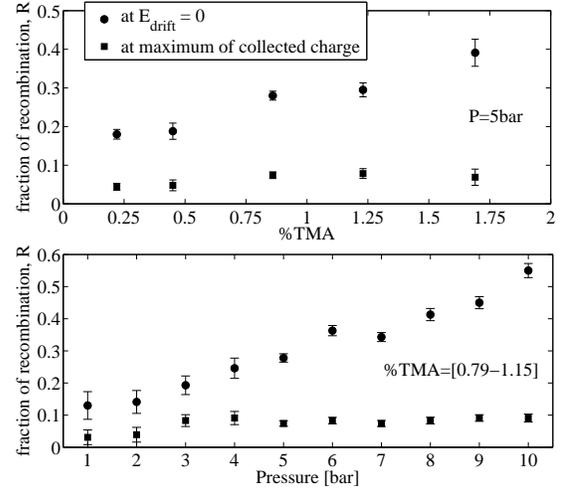}
 \caption{Fraction of charge being recombined for zero field, $1$-$Q_0/Q_{\infty}$, (circles) and for the reduced drift fields ($E_d/P$) employed during detector operation (squares).}
 \label{QQ}
 \end{figure}

Some `ad hoc' arguments given earlier in \cite{RamseyReco} and based on the Onsager \cite{Onsager} and Jaffe \cite{Jaffe} solutions, suggest that recombination in gases may be approximated by a steep function, characterized by:
\beq
\mathcal{R}(E_d/P)=(1-\frac{Q_0}{Q_\infty}) (1 - \frac{1}{1+k P/E_d})
\eeq
where $Q_{\infty}$ is the charge collected in the absence of recombination, $Q_0/Q_{\infty}$ represents the fraction of charge that escapes recombination at (near) zero field, and $k$ is an effective parameter that describes the steepness of the function. This functional dependence can be fitted to the left part of each series in Fig. \ref{SimulAndData}-down, with the help of an additional global constant that corrects for the arbitrary normalization introduced in experimental data. The fit parameter $\mathcal{R}(0)=1$-$Q_0/Q_{\infty}$ is plot in Fig. \ref{QQ} (circles), exhibiting the trend expected from recombination: high pressure and TMA concentration result in a reduction of diffusion (and in an additional increase of the ionization density in the first case) hence reducing the fraction of charge that escapes recombination at zero field. For 10~bar, e.g., this fraction does not reach 50\%. Remarkably, as argued below, a low diffusion enhances the field focusing and electron transmission, thus the net effect on the collected charge during detector operation is much less severe, going from 3\%(1bar) to 9\%(10bar) at most (Fig. \ref{QQ} squares).


Aiming at a more direct comparison, we performed a $\chi^2$ fit to the fraction of collected charge $\mathcal{F}$ through a 4-parameter model:
\beq
\mathcal{F}(E_d, E_a) = \frac{(1-\mathcal{R})\mathcal{T}}{\tn{max}[(1-\mathcal{R})\mathcal{T}]} \label{Tall}
\eeq
with $\mathcal{T}(E_d, E_a) \equiv \mathcal{T}_{MC}(E_d/g, E_a)$, and $g$ being a scaling factor that generally stayed within [0.9-1] except for 1~bar (0.81). The drift ($E_d$) and amplification ($E_a$) fields are defined as the voltage drop across the drift and amplification gap, respectively.

 \begin{figure}[h!!!]
 \centering
 \includegraphics*[width=7.5cm]{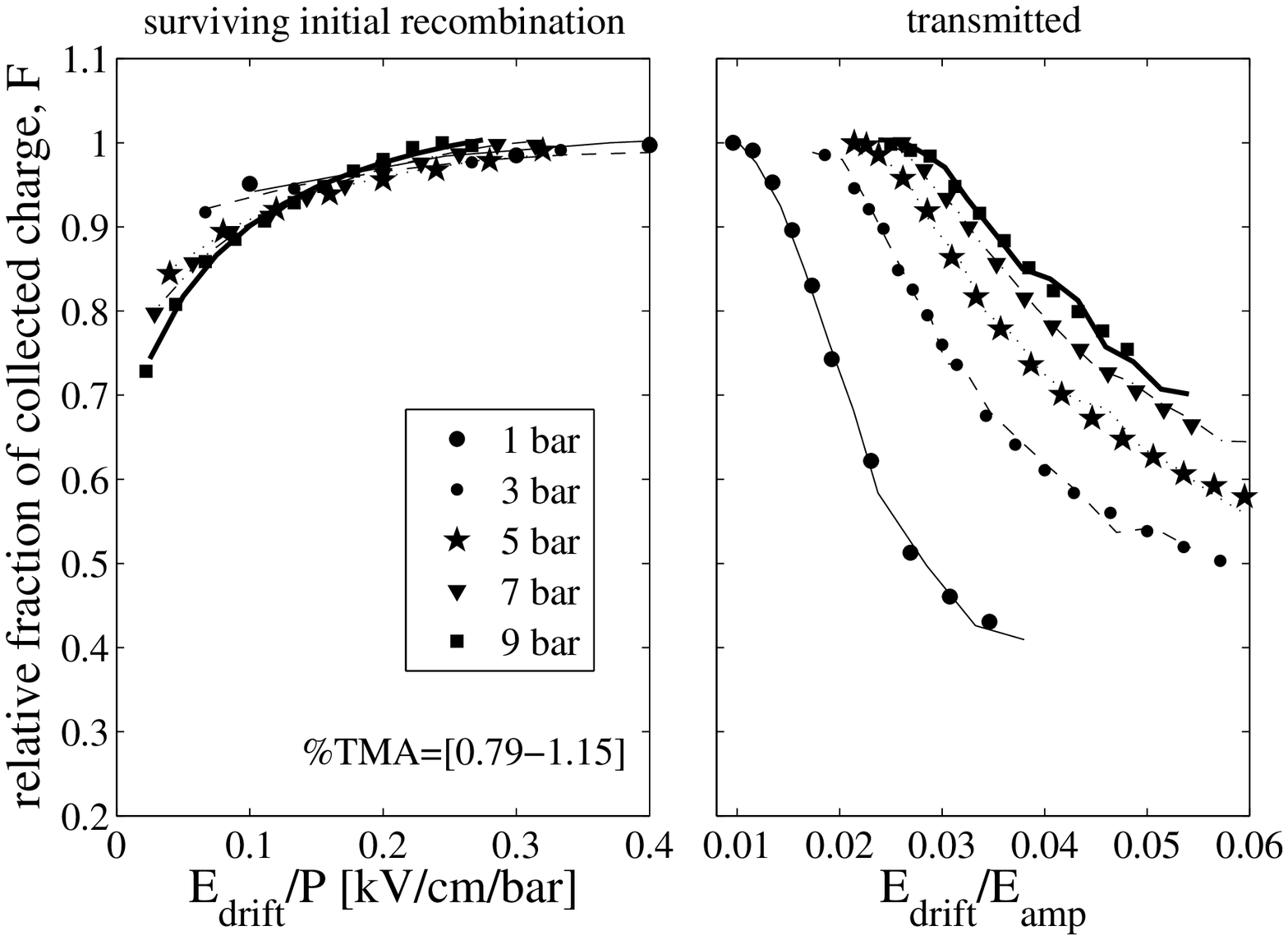}
 \includegraphics*[width=7.5cm]{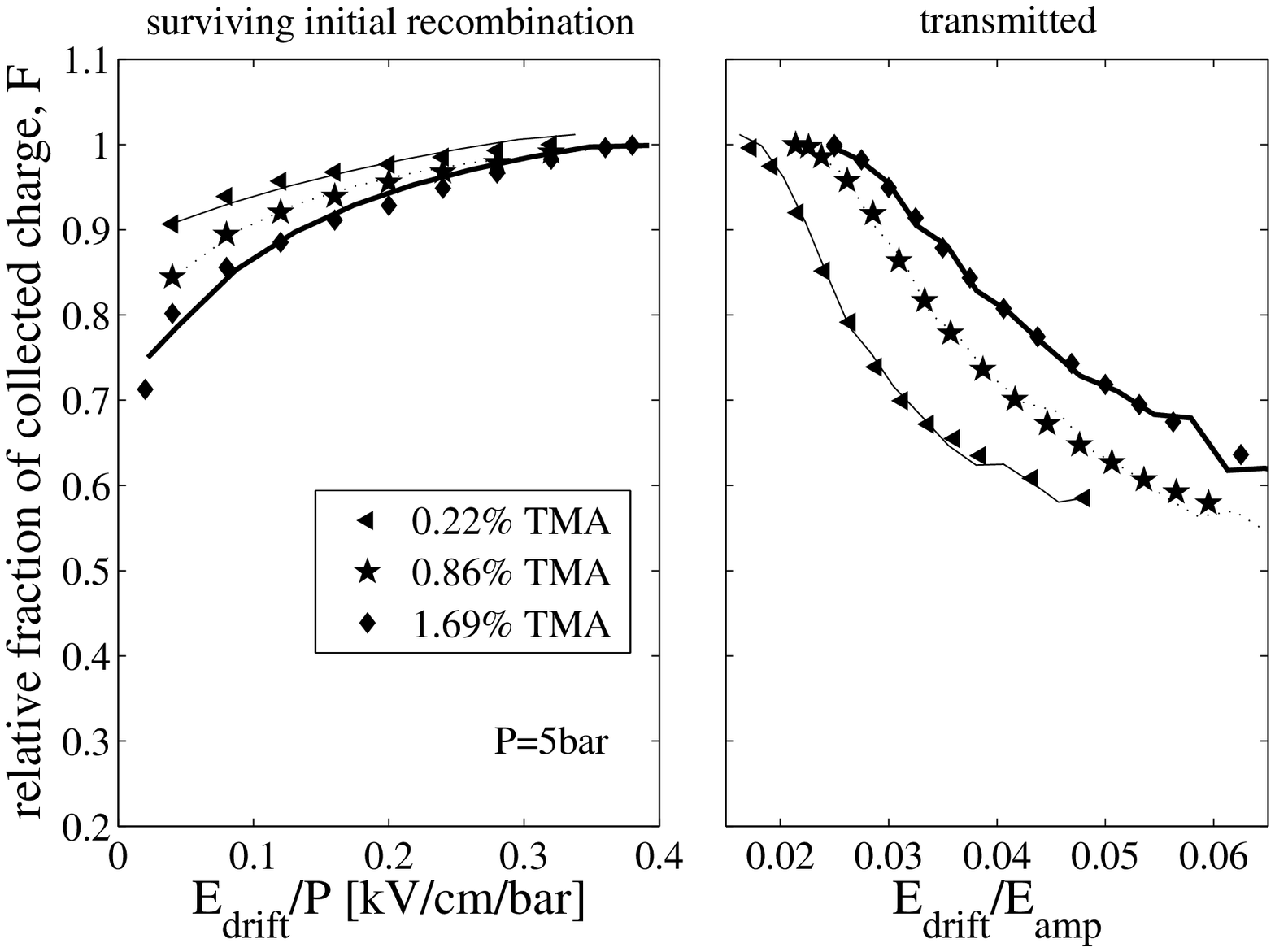}
 \caption{Relative fraction of collected charge $\mathcal{F}$ in data and for the model given in text (lines). Up: as a function of pressure, for concentrations around 1\%. Down: as a function of concentration, for 5bar.}
 \label{SimulAndDataFull}
 \end{figure}

The result of the fit is shown in Fig. \ref{SimulAndDataFull}, where the relative fraction of collected charge has been split in 2-halves. The left part shows the effect of recombination, that depends mainly on the reduced field ($E_d/P$) except for an up-down modulation following the increase-decrease of diffusion and the decrease-increase of the ionization density. On the right part, the transmission is plot as a function of the ratio of field strengths, as customary in experimental data. This representation is partly justified since in the absence of diffusion the electron transmission is largely determined by the density of electric field lines entering the holes, that is invariant upon a global scaling of both fields. In reality, and due to the presence of a gas, the right edge of the observed transmission plateaus varies within a factor $\times 3$, following the same trend as the transverse diffusion coefficient, $D_T^*$. For a typical drift field $E_d=0.2$ kV/cm/bar, simulated coefficients range from $400 {\mu}$m$/\sqrt{\tn{cm}} \times \sqrt{\tn{bar}}$ (1.5\%TMA) to $1500 {\mu}$m$/\sqrt{\tn{cm}} \times \sqrt{\tn{bar}}$ (0.1\%TMA), and these strong differences persist when the electron progresses further into the hole. Hence lower transverse diffusion leads to stronger field focusing, as intuitively expected.

The consistent behaviour of $\mathcal{F}$ and the qualitative agreement with the analysis performed in \cite{XeTMADiego}, \cite{HerreraReco} and independently in \cite{Yasu} point to the presence of charge recombination in Xe-TMA at fields $\leq 300$ V/cm/bar, increasing with pressure and with the concentration of quencher. It is however impossible, within the available measurements and simulations, to assure that the observed recombination is entirely taking place in the conversion volume. A sizeable fraction of charge recombination could be taking place at the entrance of the Micromegas holes. The increase in ionization density resulting from the increased pressure and quencher concentration can potentially have a similar impact there, leading to the recombination of the incoming electrons from the drift region with the outgoing ions produced in the preceding avalanches.


\section{Gas amplification}

The microscopic tracking of the electron trajectories including avalanche multiplication is an intense task: when performed in a standard computer it leads to computing times up to $10\tn{min} \times \tn{cpu}$ per primary electron, for the highest pressures and gains.
Nonetheless, calculations for realistic gains of several 100's as those needed for $\beta\beta0$ experiments \cite{XeTMADiego} become possible, and they were realized in the present work by occupying 10 cpu cores during approximately 3~months.

\begin{figure}[h!!!]
 \centering
 \includegraphics*[width=\linewidth]{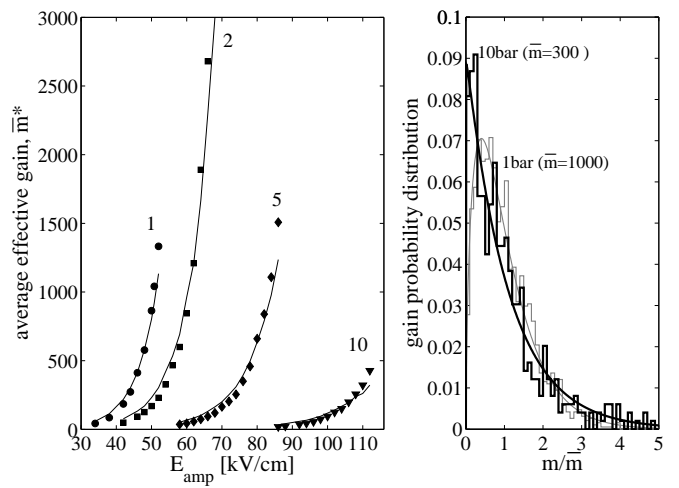}
 \caption{Left: experimental systematics of the effective gain vs amplification field with different labels indicating the operating pressure in bar. TMA concentrations are 1.22\%(1), 1.01\%(2), 1.24\%(5), 1.71\%(10) and lines indicate the simulation results. Right: simulated gain spread and fit to Polya functions.}
 \label{Gain}
 \end{figure}

Normally, in order to describe the avalanche multiplication in simulation it is necessary to determine the probability ($r_p$) of Penning transfer processes \cite{Ozkan}, that are dominated in the present case by the reaction:
\beq
\tn{Xe}^* + \tn{TMA} \rightarrow \tn{Xe} + \tn{TMA}^+ + \tn{e}^- \label{PenningEq}   ~~~~,
\eeq
Such a probability was evaluated through the following procedure: i) we started from the (comparatively simpler) estimate performed earlier under a parallel plate approximation \cite{P-F}, and obtained a reasonable description of the measured gain for any given amplification field after 3-4 simulations; ii) an improved estimate was obtained from an interpolation of the simulated $r_p$ vs gain points to the measured gain; iii) as customary, $r_p$ was assumed to be field-independent, therefore its final value and uncertainty was obtained from the statistical combination of the values obtained for the high and low field regions of each gain vs $E_a$ series. In each case we compute the effective gain as:
\beq
\bar{m}^*=\mathcal{T}(1-\mathcal{R}) \times \bar{m}
\eeq
The ($\leq10\%$) corrections due to transmission and recombination were obtained from the fits in Fig. \ref{SimulAndDataFull}. Simulated and measured gain curves for the most representative cases are compiled in Fig. \ref{Gain} for TMA concentrations around 1\% at 1bar ($r_p=0.25$), 2bar ($r_p=0.23$), 5bar ($r_p=0.12$) and 10bar ($r_p=0.14$).
Figure \ref{Gain}-right shows the distribution of gains ($m$) around the average gain $\bar{m}$, together with a fit to a Polya function in the form given in \cite{RobSta}. These distributions play an important role on the achievable energy resolution and maximum gain \cite{Zerguerras}, and they are usually characterized through their width $f=\sigma_{m/\bar{m}}^2$. The distribution of avalanche gains becomes a nearly perfect exponential at high pressure ($f=1$), since operation at smaller reduced fields $E_a/P$ is enforced \cite{Legler, Schubohm}. This fact, together with the increased charge recombination, anticipates a deterioration of the energy resolution at high pressure as demonstrated below.

\subsection{Penning transfer probability}

\begin{figure}[h!!!]
 \centering
 \includegraphics*[width=7.5cm]{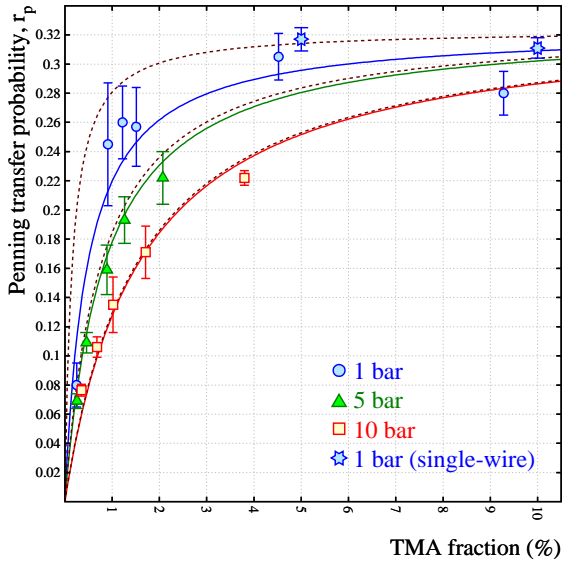}
 \caption{Penning transfer probabilities obtained from the simulation of the multiplication process for different pressures and TMA concentrations. Lines represent the simplified transfer model introduced in text, that assumes the dominance of the Xe$^*$ $^3P_1$ state. Dashed lines provide a straightforward extrapolation of the transfer probabilities from the assumed $50\mu$m amplification region to an infinite space.}
 \label{PenningAll}
 \end{figure}

The ability to capture the microscopic atom-molecule reactions leading to the effective transfer probabilities building up $r_p$ has been demonstrated in \cite{Ozkan} for single-wire and for parallel plate geometries, but only for Argon mixtures. 
It is hence reassuring to note the approximate agreement between the Penning trends obtained in this work and those coming from the modelling of the single-wire data obtained earlier by Ramsey and Agrawal in \cite{Ramsey} (Fig. \ref{PenningAll}, stars). These latter values were extracted from the measured gain curves by following the same procedure employed in \cite{Ozkan}.

We attempted a simplified description of the overall Penning systematics given in Fig. \ref{PenningAll} by considering only the processes of excimer formation $k_{ex\!c}$, Penning transfer and quenching ($k_{p}$, $k_{Q}$), together with an effective lifetime of the state undergoing Penning, $\tau^*$. The problem is further simplified by considering only the two lowest lying Xenon excited states $E_{^3\!P_1}=8.43$ eV, $E_{^3\!P_2}=8.31$ eV and neglecting Penning transfers from the Xenon excimers. The latter possibility is a highly plausible one since their energy falls well below the vertical ionization potential of TMA: $IP_{_\tn{v}}=8.44$ eV \cite{Cardoza}. Under these assumptions, the Penning transfer probability can be simply written as:
\bear
r_p = &&f_1\frac{c P k_{p,1}}{(1-c)^2P^2k_{ex\!c,1} + c P (k_{p,1} + k_{Q,1}) + 1/\tau_1^*} + \nonumber \\
      &&f_3\frac{c P k_{p,3}}{(1-c)^2P^2k_{ex\!c,3} + c P (k_{p,3} + k_{Q,3})} \label{rPform}
\eear
Sub-indexes 1, 3 refer to the singlet and triplet state, respectively, and the probability of populating each of them is labeled as $f_1$,
$f_3$.\footnote{These probabilities, if neglecting the atomic cascade, can be estimated in first order from Garfield and they approach $f_{1,3}\simeq0.5$.} Although a fit to eq. \ref{rPform} requires only 5 independent parameters, it was impossible to constrain $k_{p,3}$ from the available data, its value representing unfortunately an important contribution for the extrapolation to low concentrations. It is sensible, however, to consider only the energy transfers from the singlet state. Its energy is nowadays known to be compatible with the vertical ionization potential of TMA within 10 meV. This is a sub-thermal energy difference at ambient temperature and it represents a remarkable fine-tuning of Nature, supporting the speculation that a resonant transfer may be taking place. After this, only 3 independent parameters are left in the fit:
\beq
r_p = \frac{c P}{(1-c)^2P^2a_{1} + c P a_{2} + a_3} \label{PenningEq}
\eeq
The result of the fit is shown in Fig. \ref{PenningAll} (continuous lines), with parameters:
\bear
& a_1 & = 0.0047 \pm 0.0004~\tn{bar}^{-1}\\
& a_2 & = 3.1    \pm 0.1 \\
& a_3 & = 0.010  \pm 0.004~\tn{bar}
\eear

By making use of the known transfer rate of the singlet state $k_{ex\!c,1}= 51 ~ 10^6$ bar$^{-2}$ s$^{-1}$ \cite{kexc}:
\beq
\tau_1^* = \frac{a_{1}}{a_{3}} \frac{1}{k_{ex\!c,1}} = 9.2 \pm 2.5 \tn{ns}
\eeq
Due to the strongly model-dependent analysis procedure, the approximate agreement with the life-time of the singlet state $\tau_1=4.3\pm0.5$ns \cite{singletLife} must be taken with care. Effective life-times of atomic states can increase easily by 3 orders of magnitude for cm-scale gas cells due to photon-trapping \cite{singletLife,Holstein}, although the reduced nature of the amplification hole ($50\mu$m) seems compatible with the small enhancement observed for $\tau_1^*$ in the present situation. In Fig. \ref{PenningAll} the result corresponding to $\tau_1^*\!\rightarrow\!\infty$ has been overlaid, a situation that would correspond to the Penning transfer rates observable in an unconfined space (e.g. a typical drift/conversion region).

\subsection{Energy resolution for 22~keV X-rays}

Each primary X-ray releases about $n_e=\varepsilon/W_I = 900$ primary electrons, given an average energy for creating an electron-ion pair of $W_I=24.8$ eV (in pure Xenon). In the range of TMA concentrations studied here we consider $W_I$ to be the same for both species, an assumption that should lead to an error on the estimate of the direct ionization of $\leq 1\%$.

\begin{figure}[h!!!]
 \centering
 \includegraphics*[width=7.5cm]{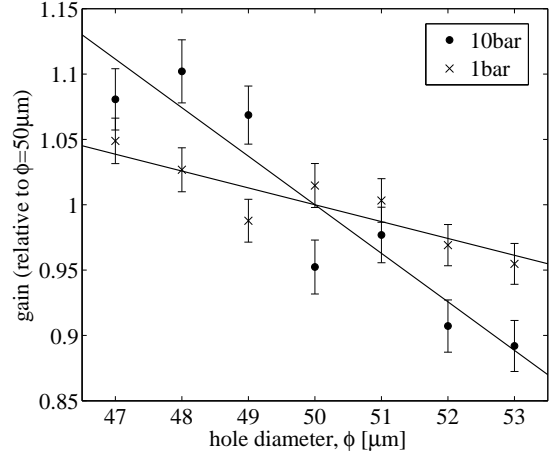}
 \caption{Gain as a function of the diameter size, relative to the reference geometry ($\phi=50\mu$m). Normalization obtained from the super-imposed linear fits.}
 \label{GainSlope}
 \end{figure}

A priori, the energy resolution in present conditions can be expected to be dominated by the stochastic nature of the avalanche multiplication process $f$ (Fig. \ref{Gain}-right), i.e. $\mathfrak{R} \sim 2.35 \sqrt{fW_I/\varepsilon}$. This is because, as suggested by \cite{RobSta}, during the operation of a gaseous detector in practical conditions this contribution always exceeds the intrinsic fluctuations of the initial number of electrons, determined by the Fano factor (e.g., $F=0.15\pm0.02$ in pure Xenon \cite{Davebb0}). In detail, a complete description requires yet additional fluctuations related to charge losses ($(1-\mathcal{R}), \mathcal{T}$), signal to noise (S/N) and mechanical tolerances to be considered (see \cite{XeTMADiego}, for instance):

\bear
& \mathfrak{R} & = 2.35 \sqrt{\sigma_{int}^2 + \sigma_{mech}^2 + \sigma_{S/N}^2} \label{ResForm}\\
& \sigma_{int} &  = \sqrt{F + f + \mathcal{T}\mathcal{R} + (1-\mathcal{T})}\frac{1}{\sqrt{n_{e}}}  \\
& \sigma_{mech} & = \left|\frac{1}{\bar{m}}\frac{dm}{d\phi}\right| \sigma_\phi \\
& \sigma_{S/N}  & = \frac{ENC}{\bar{m}}\frac{1}{n_{e}} \\
& n_e & = (1-\mathcal{R})\mathcal{T}\frac{\varepsilon}{W_I} \label{intrins}
\eear
where ENC is the equivalent noise charge at the amplifier input, and all other magnitudes have been defined.
The impact that the variations of the hole size, characterized through $\sigma_{mech}$, have on the energy resolution (either those stemming from the deviations from an ideal circle or hole-to-hole size variations) is difficult to assess, so the proposed idealized treatment must be necessarily considered as approximate.
The estimate of $\sigma_{mech}$ may be characterized by the magnitude $\frac{1}{\bar{m}}\frac{dm}{d\phi}$ and a typical diameter spread $\sigma_\phi$, under the assumption that the cloud of 900 primary electrons is large enough to cover a statistically representative region. Fig. \ref{GainSlope} shows the gain dependence with the hole size. For small holes the field approaches more closely the parallel plate limit, leading to an increase in gain. More importantly, the slope is nearly a factor $\times 3$ higher at high pressure (table \ref{TableSlope}).



\begin{table}[h]
  \centering
  \begin{tabular}{|c|c|c|}
     \hline
     P[bar] & gain & $-\frac{1}{\bar{m}}\frac{dm}{d\phi} [\%/\mu\tn{m}]$ \\
     \hline
     1   & 350  & $1.1\pm0.3$ \\
     1   & 1350 & $1.3\pm0.4$ \\
     10  & 100  & $3.0\pm0.6$ \\
     10  & 150  & $3.7\pm0.8$ \\
     \hline
   \end{tabular}
  \caption{Relative gain variation with respect to the hole diameter for different pressures and gains.}\label{TableSlope}
\end{table}

For operation close to the edge of the electron transmission plateau it was observed in simulation that the concurrent increase of transmission with the diameter size can lead to an approximately compensated situation for the effective gain $\bar{m}^*$, representing a limited analogy with the classical compensation in standard Micromegas \cite{YannisComp}. In this discussion we assume for simplicity that measurements were performed far enough from the transmission edge so that only amplification is modified by the hole diameter spread.

\begin{figure}[h!!!]
 \centering
 \includegraphics*[width=7cm]{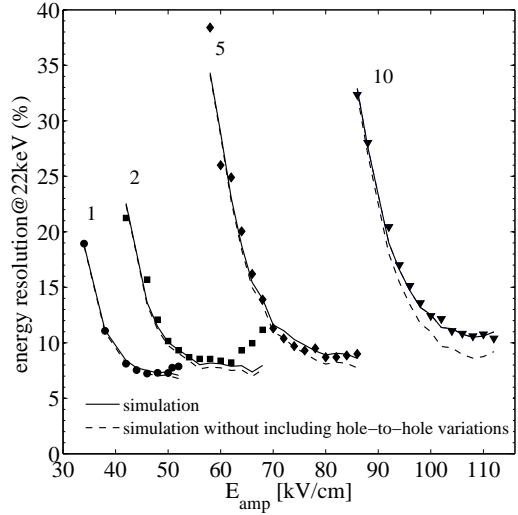}
 \caption{Energy resolution for different TMA concentrations and pressures 1.22\%(1bar), 1.01\%(2bar), 1.24\%(5bar), 1.71\%(10bar). Continuous lines represent the results obtained with the full simulation described in text, while for the dashed lines the contribution coming from the variations of the holes' diameter has been eliminated.}
 \label{Res}
 \end{figure}

Energy resolution measurements are shown in Fig. \ref{Res}, together with simulation results after including all identified contributions (continuous lines). The ENC has been adjusted to describe the low-field (low-gain) region, providing values in the range 2000-4000e$^-$; a typical diameter spread of $\sigma_\phi=0.6 \mu$m was assumed, with the remaining contributions being taken directly from simulation and formula \ref{ResForm}. As shown, the contribution of the hole accuracy is strongly hinted by data under an assumption on the diameter spread that seems realistic. Indeed, the value for $\sigma_{mech}$ indirectly derived for the large demonstrator in \cite{XeTMADiego} is consistent with the present analysis, both for 1 bar and 10 bar data, if setting $\sigma_\phi=1 \mu$m. The presence of topological domains with different average values of $\phi$ (within the quoted $1\mu$m) due to the larger areas considered there ($8$mm$ \times 8$mm per pixel), could naturally explain the additional spread.

The deterioration observed at high gains for $P=2$bar points to the presence of feedback, an effect that can completely dominate the detector response (most strikingly for operation under pure Xenon, \cite{Balan}, where the observed resolutions exceed by a factor $\times 4$ the ones obtained here). For the type of readout and gas mixture studied in this work, the energy resolution plateau for which sufficient S/N and low photon-feedback can be simultaneously achieved extends along a comfortable field range of about 10-20\%.

\begin{figure}[h!!!]
 \centering
 \includegraphics*[width=6.6cm]{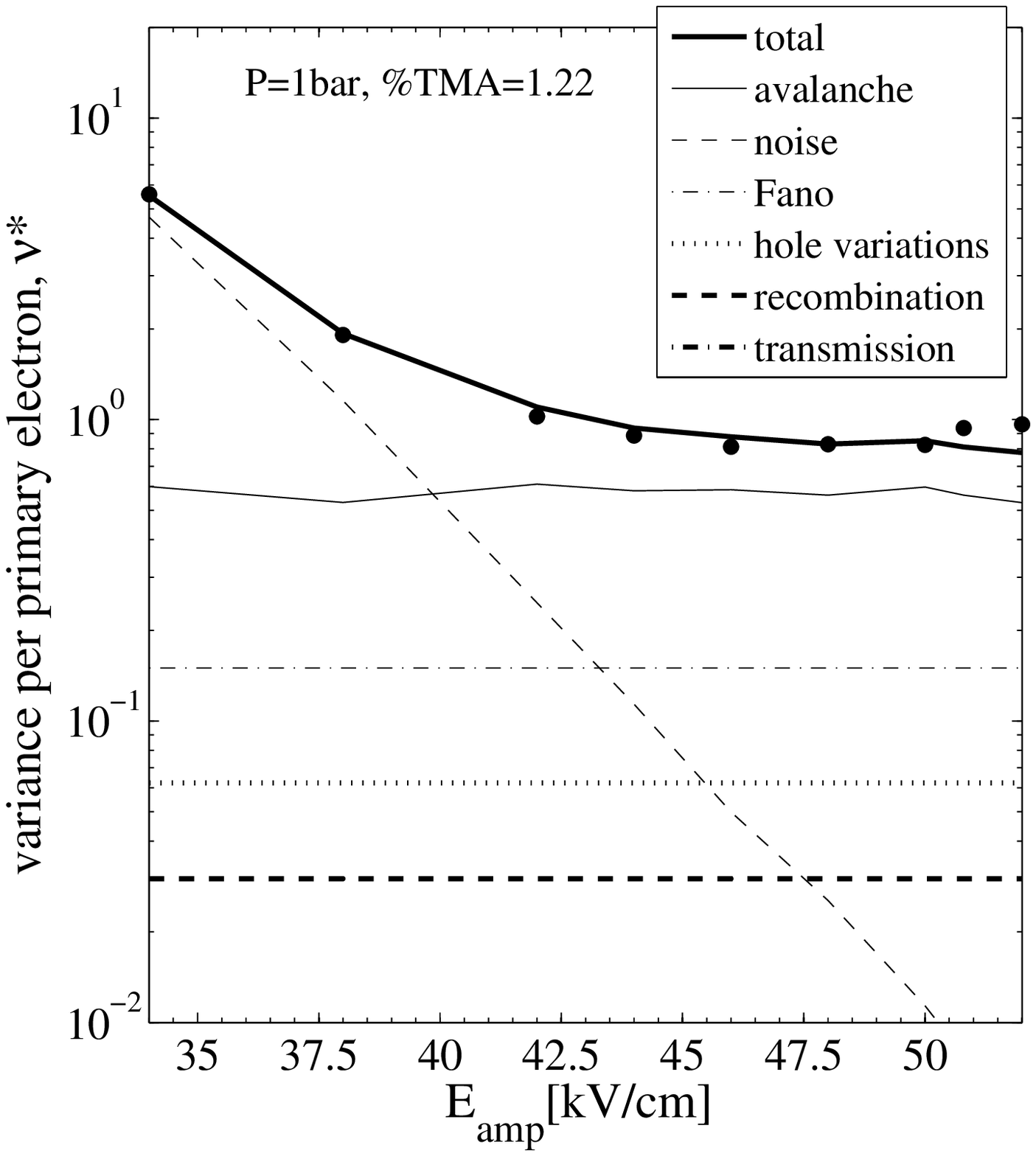}
 \includegraphics*[width=6.6cm]{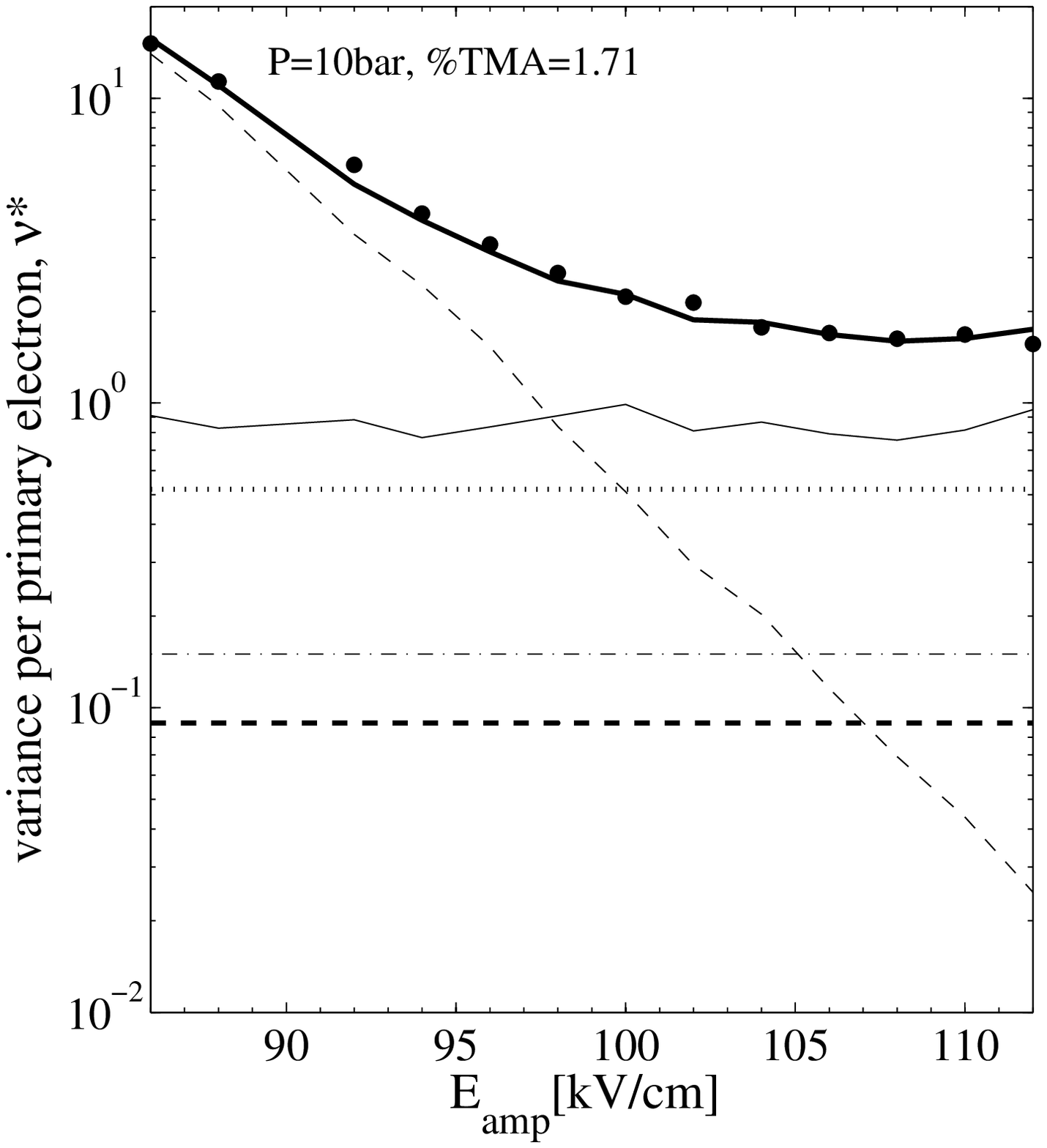}
 \includegraphics*[width=6.6cm]{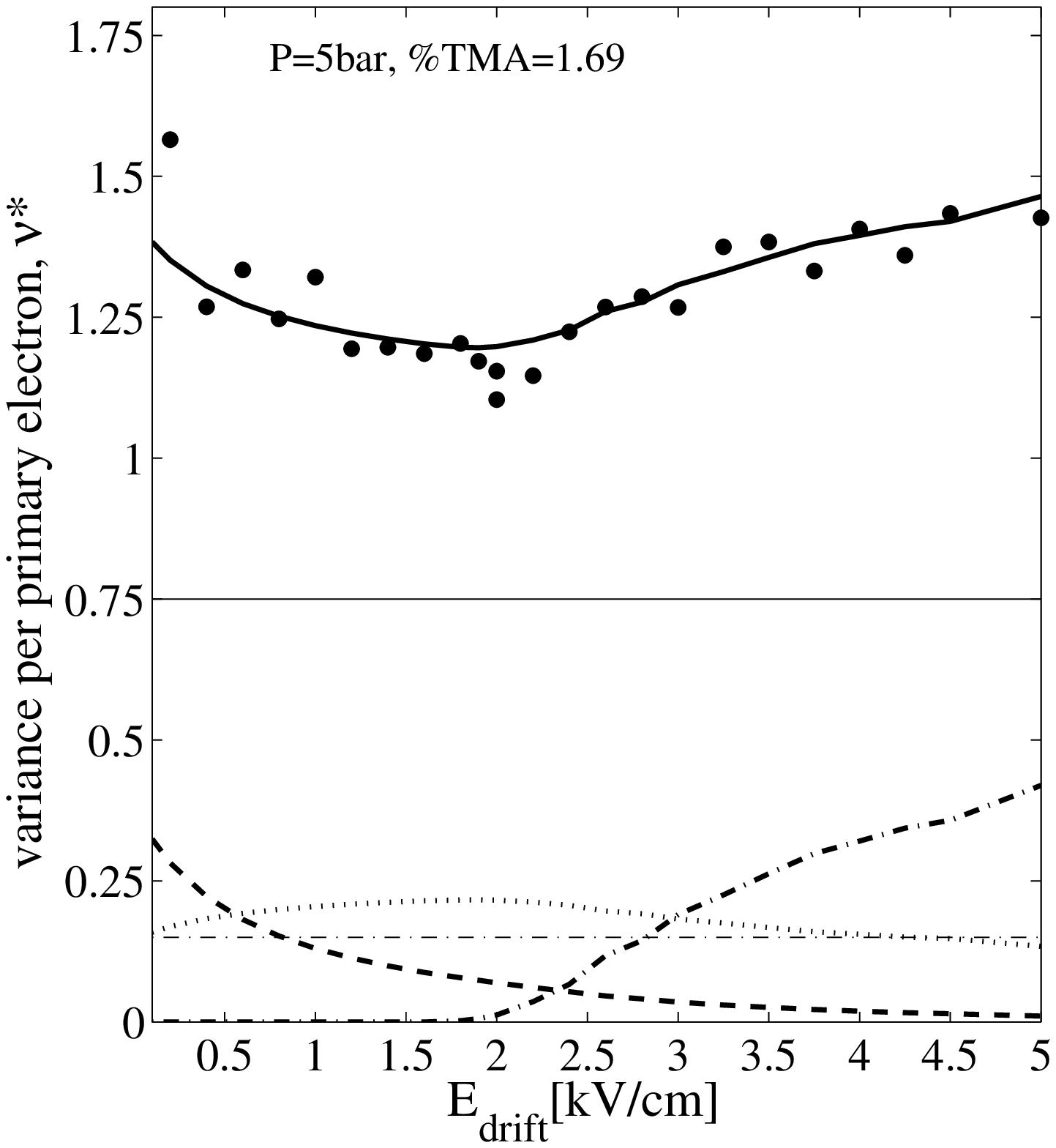}
 \caption{Variance per electron, $v^*$, and the various contributions discussed in this work. Up: gain scan (1bar). Middle: gain scan (10bar). Down: drift field scan (5bar).}
 \label{Var}
 \end{figure}

The influence of each contribution to the energy resolution can be more easily extracted from the effective variance per electron,
defined as:
\beq
v^* = \left(\frac{\mathfrak{R}}{2.35} \right)^2  n_{e} \label{refvstar}
\eeq
and that is shown in Fig. \ref{Var} for three representative field scans.

\section{Discussion on Fano factor and beyond-intrinsic energy resolution in Xenon}

In the presence of Penning transfer reactions, the energy resolution is modified as:
\bear
& \mathfrak{R}_p & = \sqrt{\frac{n_{e}}{n_{e,p}}\mathfrak{R}^2 - 2.35^2 \frac{(F - F_p)}{n_{e,p}}}\label{ref} \\
&\frac{n_{e,p}}{n_{e}} & = 1+\frac{N_{ex}}{N_{I}}r_p(E_d)
\eear
It depends on the number of excitations (susceptible of Penning transfer) relative to the number of ionizations that are generated by the primary particle, $\frac{N_{ex}}{N_{I}}$, on the Fano factor of the Penning mixture, $F_p$, and on the transfer probability at the typical drift fields, $r_p(E_d)$. The determination of $F_p$ is obscured in the present case by the hypothesized contribution $\sigma_{mech}$, that exceeds the naive upper value set by $F_p=F$ for nearly all pressures. At 1 bar and 1\%TMA admixture, however, and in the range $E_a=45$-$50$ kV/cm the Fano factor expectedly becomes the second contribution in importance just after the avalanche fluctuations (Fig. \ref{Var}-up), while the Penning transfer becomes maximal (Fig. \ref{PenningAll}). In such conditions, a determination can be attempted.

In order to extract $F_p$, a set of additional assumptions is needed:
first we assume that the Penning transfer probability extracted from the modelling of the amplification process is similar to the one
at typical drift fields, so $r_p(E_d) \equiv r_p(E_a) \equiv r_p = 0.24 \pm 0.04$.
Additionally, we assume that the fraction of excited states susceptible of Penning transfer is directly related to the number of scintillation photons, under which $\frac{N_{ex}}{N_{I}} = \frac{W_I}{W_{sc}}$, with W$_{sc}$ being the energy that it takes to create a photon in Xenon, $W_{sc}=60\pm20$ eV \cite{Josh, Joaquim, Michel}:
\beq
\frac{n_{e,p}}{n_{e}} = 1+\frac{W_I}{W_{sc}}r_p = 1.10\pm 0.04 \label{np}
\eeq
If it is true (as speculated in the previous section) that the singlet Xe$^*$ state predominantly contributes to the transfer, $\frac{N_{ex}}{N_{I}}$ becomes smaller and the ratio $\frac{n_{e,p}}{n_{e}}$ approaches unity. We will assume for simplicity that this situation is approximately comprised within the assigned uncertainty.
A comparison with formulas \ref{ResForm} and \ref{refvstar} in the region $E_a=45$-$50$kV/cm leads finally to:
\bear
F_p && = \frac{n_{e,p}}{n_{e}}v^* - f - \mathcal{T}\mathcal{R} - (1-\mathcal{T}) - \sigma_{mech} - \sigma_{S/N} = \nonumber \\
    && = 0.20\pm0.06 \label{Fp}
\eear
The evaluation has been performed for $1\tn{bar}$ and $1\%$ TMA and makes use of the experimental measurement of $v^*$
together with the parameters estimated in the Micromegas model developed in
previous sections. The quoted uncertainty is mainly of systematic origin, dominated by the estimate of $\sigma_{mech}$.

The Fano factor determined through eq. \ref{Fp} is compatible with the one for pure Xenon, and nearly compatible with the minimum expected
value \cite{Alkhazov}:
\beq
F_{p,min} = F(1-r_p) = 0.11\pm0.02
\eeq
The intrinsic energy resolution of the gas mixture is thus:
\beq
\mathfrak{R}_{0,Xe-TMA}\! =\! 2.35 \sqrt{\frac{F_p}{n_{e,p}}}\! = \! (0.50\% \pm 0.08\%)\! \sqrt{\frac{1 \tn{MeV}}{\varepsilon}}
\eeq
to be compared with the one expected in pure Xe \cite{Davebb0}:
\beq
\mathfrak{R}_{0,Xe}  = 2.35\sqrt{\frac{F}{n_e}} = (0.45\% \pm 0.03\%) \sqrt{\frac{1 \tn{MeV}}{\varepsilon}}
\eeq
This intrinsic energy resolution of the mixture (labeled with subscript $0$) is important in scenarios where the multiplication process is not the limiting factor in the reconstructed energy (e.g. \cite{NEXT1}, \cite{Jochen}). From such a perspective, it is implied by this analysis that the energy resolution achievable in Xe-TMA admixtures will represent only a modest improvement over pure Xenon, if at all. Since the assumed Penning transfer probability is already close to the asymptotic value $r_p\simeq0.30$ (eq. \ref{PenningEq}), a strong improvement seems unlikely for virtually any Xe-TMA gas admixture.

The analysis presented suggests that the quality of microbulk Micromegas and microscopic modelling has reached a situation where the Fano factor of a gas mixture and its intrinsic energy resolution may be determined from the energy resolution obtained with X-rays (at least for certain operating conditions). A more systematic experimental verification of the indirect procedure here introduced is still needed, however, normally involving the use of different wafers or positions within the same wafer. On the other hand, the steady growth of the electron counting techniques performed on similar amplification structures can provide the necessary complementary path \cite{Jochen}.

\section{Conclusions}

We have shown that the behaviour of Micromegas manufactured in the microbulk and operated under Xenon/trimethylamine mixtures is well described by state of the art microscopic simulations in a broad range of concentrations and pressures. When irradiated with 22 keV X-rays, the fraction of charge arriving at the holes can be interpreted as the result of electron-ion recombination at low drift fields, and the loss of field-focusing at high drift fields (modulated by the transverse diffusion). A proper description of the gas amplification requires the presence of Penning transfers, expected due to the proximity between the ionization potential of trimethylamine and the energy levels of the excited states in Xenon. The probability of such processes, reaching maximum values in the range $r_p=20$-$30$\%, can be interpreted with a simple model that involves the resonant energy transfer from Xenon atoms in the $^3P_1$ state to trimethylamine molecules.

The energy resolution shows a degradation at high pressure slightly (but systematically) beyond the expected contributions, and it has been shown to be compatible with a higher sensitivity of the amplification process to the mechanical tolerances in such conditions. In that respect, the accuracy of the hole manufacturing process has been evaluated to be in the range $\sigma_\phi \leq 0.6 ~\mu$m (in a 1 mm$^2$ region) to $1~\mu$m (in a 100 mm$^2$ region), values beyond which the simulation becomes incompatible with the measurements.

At low pressure, where the Penning transfers are higher and the influence of the mechanical accuracy was estimated to be negligible, the present work yields a Fano factor in Xe/TMA of $F_p(1\tn{bar}, 1\%\tn{TMA}) = 0.20\pm0.06$, together with an increase of the primary ionization by $10\% \pm 4\%$ relative to pure Xenon. The intrinsic resolution of the Xe-TMA gas mixture amounts to:

$\mathfrak{R}_{0,Xe-TMA}\! = \! (0.50\% \pm 0.08\%)\! \times \sqrt{1 \tn{MeV}/\varepsilon}$.

\ack
DGD acknowledges the support of Steve Biagi, Carlos Oliveira and Josh Renner (who provided the Gmsh/Elmer Garfield++ interface),
Yasuhiro Nakajima, and (specially) Azriel Goldschmidt for his continuous support and interest on the `black magic' involved on gaseous electronics. We want to thank our NEXT colleagues for providing an always stimulating atmosphere.

 The excellent technical support and service from the CERN detector workshop
and Servicio General de Apoyo a la Investigaci´on-SAI of the University of Zaragoza
was essential to our investigation.
This work was funded by  Spanish Ministry of Economy and Competitiveness (MINECO)
under grants Consolider-Ingenio 2010 CSD2008-0037 (CUP) and CSD2007-00042 (CPAN),
FPA2011-24058, FPA2013-41085, FPA2008-03456, FPA2009-13697, and
by the T-REX Starting Grant ERC-2009-StG-240054 of the
IDEAS program of the 7th EU Framework Program. Part of these grants are funded
by the European Regional Development Plan (ERDF/FEDER).

%


\begin{thebibliography}{00}
\bibitem{muBulk} S. Andriamonje et al., JINST 5(2010)P02001.
\bibitem{MMyannis} Y. Giomataris, Ph. Rebourgeard, J. P. Robert, G. Charpak, Nucl. Instr. Meth. A, 376(1996)29.
\bibitem{PacoMeas} F. J. Iguaz et al., JINST 7(2012)P04007.
\bibitem{XeTMADiana}  S. Cebrian et al., JINST 8 (2013) P01012.
\bibitem{Zerguerras} T. Zerguerras et al., Nucl. Instr. Meth. A, 772(2015)76.
\bibitem{RobSta}  H. Schindler, S. F. Biagi, R. Veenhof, Nucl. Instr. Meth., A 624(2010)78.
\bibitem{XeTMADiego}  D. Gonzalez-Diaz et al. (NEXT Collaboration), arXiv: 1504.03678 [physics.ins-det], submitted to Nucl. Instr. Meth. A.
\bibitem{XaviRadio} S. Aune et al., JINST 8 (2013) C12042.
\bibitem{HectorRadio}  T. Dafni et al., Astropart. Phys. 34(2011)354.
\bibitem{nTOF} N. Colonna et al., Nucl. Instr. Meth. B, 269(2011)3251.
\bibitem{IAXO} E. Armengaud et al., JINST 9(2014)T05002.
\bibitem{TREX} F. J. Iguaz et al., arXiv:1503.07085 [physics.ins-det].
\bibitem{HPXeChina} Xiandong Ji, private communication.
\bibitem{Gotthard} R. Luescher et al., Phys. Lett. B, 434(1998)407.
\bibitem{NEXT1} \'Alvarez~V. et al. (NEXT Collaboration), JINST 7(2012)T06001.
\bibitem{NEXT2} D. Lorca et al., (NEXT Collaboration), JINST 9(2014)P10007.
\bibitem{YannisComp} Y. Giomataris, Nucl. Instr. Meth. A, 419(1998)239.
\bibitem{NarrowGap} D. Atti\'e et al., JINST 9(2014)C04013.
\bibitem{HerreraReco} D. C. Herrera et al., PoS(TIPP2014)057.
\bibitem{Balan} C. Balan et al. JINST 6(2011)P02006.
\bibitem{Garfield} Garfield++ – simulation of tracking detectors, http://cern.ch/garfieldpp (accessed 8 May, 2015).
\bibitem{Elmer} Elmer: Open Source Finite Element Software for Multiphysical
Problems, http://www.csc.fi/english/pages/elmer (accessed 8 May, 2015).
\bibitem{gmesh} C. Geuzaine and J.-F. Remacle, Int. J. Numer. Meth. Eng. 79(2009)1309.
\bibitem{Dave} D. R. Nygren, J. Phys. Conf. Ser. 460(2013)012006.
\bibitem{Jin} Jin Li, arXiv:1503.07320 [hep-ph].
\bibitem{DiegoProc} Diego Gonzalez-Diaz et al., arXiv:1504.03640 [physics.ins-det].
\bibitem{Magboltz} S. F. Biagi, Nucl. Instrum. Meth. A 421(1999)234.
\bibitem{RobTransp} K. Nikolopoulos, P. Bhattacharya, V. Chernyatin, R. Veenhof, JINST, 6(2011)P06011.
\bibitem{RobTrans} P. M. M. Correia et al., JINST 9 (2014) P07025.
\bibitem{XeTMADiegoOLD} V. Alvarez et al. (NEXT Collaboration), JINST 9(2014)C04015.
\bibitem{Yasu} Y. Nakajima et al., arXiv:1505.03585 [physics.ins-det].
\bibitem{RamseyReco} A. Bolotnikov, B. D. Ramsey, Nucl. Instr. Meth., A 428(1999)391.
\bibitem{Onsager} L. Onsager, Phys. Rev. 54(1938)554.
\bibitem{Jaffe} G. Jaffe, Ann. Phys. 42(1913)303.
\bibitem{P-F} Diego Gonzalez-Diaz et al., arXiv:1504.03640 [physics.ins-det], submitted to Nucl. Instr. Meth. A.
\bibitem{Legler} W. Legler, Z. Phys. 140(1955)221.
\bibitem{Schubohm} H. Schlumbohm, Z. Phys. 151(1958)563.
\bibitem{Ozkan} O. Sahin et al., JINST 5(2010)P05002.
\bibitem{Ramsey} B. D. Ramsey, P. C. Agrawal, Nucl. Instr. Meth. A, 278(1989)576.
\bibitem{Cardoza} Job D. Cardoza et al., J. Phys. Chem. A, 112(2008)10736.
\bibitem{kexc} J. Galy, K. Aouame, A. Birot, H. Brunet, P. Millet, J. Phys. B, At. Mol. Opt. Phys. 26(1993)477.
\bibitem{singletLife} Y. Salamero, A. Birot, H. Brunet, J. Galy, P. Millet, J. Chem. Phys. 80(1984)4774.
\bibitem{Holstein} T. Holstein, Phys. Rev. 83(1951)6.
\bibitem{Davebb0} David Nygren, Nucl. Instr. Meth., A 337(2009)603.
\bibitem{Josh} J. Renner et al. (NEXT collaboration), Nucl. Instr. Meth. A, 793(2015)62.
\bibitem{Joaquim} L. M. P. Fernandes et al., JINST 5(2010)P09006.
\bibitem{Michel} L. Serra, M. Sorel, et al. (NEXT collaboration), JINST 10(2015)P03025.
\bibitem{Alkhazov} G. D. Alkhazov, A. P. Komar, A. A. Vorobev, Nucl. Instr. Meth. 48(1967)1.
\bibitem{Jochen} C. Krieger et al., Nucl. Instr. Meth. A, 729(2013)905.
\end{thebibliography}
\end{document}